\documentclass[10pt]{article}

\usepackage{amsmath}
\usepackage{amssymb}
\usepackage{epsf}

\allowdisplaybreaks

\def\half{\frac{1}{2}}

\def\tp{\; {}^{t}\hspace{-1pt}} 

\def\CN{{\mathcal N}}
\def\CW{{\mathcal W}}
\def\CO{{\mathcal O}}

\def\bN{{\bf N}}

\def\BZ{{\mathbb Z}}
\def\BR{{\mathbb R}}
\def\BC{{\mathbb C}}

\def\bmat{\begin{matrix}}
\def\emat{\end{matrix}}
\def\bpmat{\begin{pmatrix}}
\def\epmat{\end{pmatrix}}

\def\tr{{\rm tr}\,}
\def\Det{{\rm Det}\,}
\def\det{{\rm det}\,}
\def\Pf{{\rm Pf}\,}

\def\diag{{\rm diag}\,}
\def\bdiag{{\rm block.diag}\,}

\def\del{\partial}

\def\Asym{\raisebox{-3.6pt}{$\square$} \hspace{-7.8pt}
  \raisebox{2.8pt}{$\square$} }

\begin{document}

\baselineskip=18.2pt plus 0.2pt minus 0.1pt

\makeatletter
\@addtoreset{equation}{section}
\renewcommand{\theequation}{\thesection.\arabic{equation}}
\renewcommand{\thefootnote}{\fnsymbol{footnote}}

\begin{titlepage}
\title{
\hfill\parbox{4cm}
{\normalsize KUNS-1742\\{\tt hep-th/0110209}}\\
\vspace{1cm}
{\Large\bf
Comments on orientifold projection in the conifold \\
and $SO\times USp$ duality cascade}
}
\author{
Shin'ichi {\sc Imai}\thanks{{\tt imai@gauge.scphys.kyoto-u.ac.jp}}
{}\hspace*{5pt} and \hspace*{5pt}
Takashi {\sc Yokono}\thanks{{\tt yokono@gauge.scphys.kyoto-u.ac.jp}}
\\[7pt]
{\it Department of Physics, Kyoto University, Kyoto 606-8502, Japan}
\\[10pt]
}
\date{\normalsize October, 2001}
\maketitle
\thispagestyle{empty}
\begin{abstract}
\normalsize\noindent
We study the O3-plane in the conifold.
On the D3-brane world-volume we obtain $SO \times USp$ gauge
theory that exhibits a duality cascade phenomenon.
The orientifold projection is determined 
on the type IIB string side, and corresponds to that of O4-plane on the
dual type IIA side.
We show that SUGRA solutions of 
Klebanov-Tseytlin and Klebanov-Strassler survive
under the projection.
We also investigate the orientifold projection in the generalized
conifolds, and verify desired features of the O4-projection in the type IIA picture.
\end{abstract}

\end{titlepage}


\tableofcontents
\section{Introduction}
\label{sec:intro}

 In the past years, an extension of AdS/CFT correspondence 
\cite{M,GKP,W} has been investigated away form conformality. 
Especially, type IIB SUGRA solutions that describe 
D3-branes at the conifold singularity beautifully reproduce 
phenomena of field theories,
such as RG flow, duality cascade, chiral symmetry breaking and confinement.

 When $N$ D3-branes are placed at the conifold singularity, $\CN=1$ superconformal field
 theory which is $SU(N)\times SU(N)$ gauge theory with $2
 (\bN,\overline{\bN}) \oplus 2 (\overline{\bN},\bN)$ is realized on the
 branes \cite{KW}.
 The addition of $M$ fractional D3-branes changes the gauge groups to
 $SU(N+M)\times SU(N)$ and breaks the conformal invariance.
As we flow to IR, the gauge coupling constant of $SU(N+M)$
 diverges and Seiberg duality must 
 be performed for better description of the field theory. 
 As the dual theory has similar
 gauge groups $SU(N-M)\times SU(N)$ and matter content as the original
 theory, this process repeats successively. This is called ``RG cascade'' 
 or ``duality cascade'' \cite{KS}. 
At the bottom of this cascade, Affleck-Dine-Seiberg superpotential is  dynamically
generated \cite{ADS}. The moduli space of vacua is deformed and chiral
symmetry is broken by gaugino condensation.
The type IIB SUGRA solution of Klebanov-Tseytlin (KT solution)
 \cite{KN,KT} descibes D3-branes at the conifold singurality and incorporates this cascade. 
The NS-NS $B$ field that corresponds to the gauge couplings 
$1/g_1^2 - 1/g_2^2$ has logarithmic radial dependence.
And 5-form fluxes which corresponds the rank of the
 gauge group suitably decrease. 
 The SUGRA solution found by Klebanov-Strassler (KS solution)
 \cite{KS} furthermore reproduces far IR phenomena as well as duality cascade. 
It has asymptotically the same form as Klebanov-Tseytlin
 \cite{KT} solution, while near the origin, the singularity of the conifold is deformed
and the branes are replaced with fluxes. So it signals confinement
in the gauge theory \cite{KS,AMV}.

 In this paper, we extend these results to the $SO \times USp$ gauge theory.
In the type IIA brane configurations,
there are two possibilities to obtain $SO$ or $USp$ gauge group.
One is with O6-planes \cite{Uranga}.
Another is  with an O4-plane \cite{Lopez,EJS}.
Taking T-duality to the conifold with D3-branes, we have brane configurations with a
NS5-brane along $012345$, a NS5'-brane along $012389$ and D4-branes along
$01236$ \cite{DM}.
To obtain the $SO \times USp$ gauge groups, only the O4-plane along
$01236$ is allowed in the case.
We consider the corresponding orientifold projection in type IIB
theory.
Such projection is also discussed in \cite{ANS}.
But we give another projection by studying symmetries of the type IIB
conifold. Our projection gives the correct field theory.
Other models with O6-planes have been well studied and 
corresponding orientifold projections in type 
IIB theory are given in \cite{BI, PU, PRU2, PRU, NSW}.
 We also comment on KT/KS solutions. 
They still solve equations of motion under the projection.
Moreover we generalize the projection in the conifold to one
in the generalized conifolds.
In the type IIA picture, we have some NS5-branes and NS5'-branes
with the O4-plane.
The orientifold projection is consistent with the feature of the
O4-plane such that the gauge groups must be $(SO \times USp)^n$ \cite{Lopez,EJS}
and the total number of NS5 and NS5'-branes requires to be even \cite{Lopez,EJS,Hori}.

This paper is organized as follows. 
In section \ref{sec:Expectation}, as a heuristic step, 
we analyze type IIA brane configurations. In section
\ref{sec:DetOProj}, we determine the orientifold projection 
in gauge theory language. Then, we analyze the field theory and observe
similar phenomena as in the $SU \times SU$ case.
  In section \ref{sec:IIBSUGRA},
we give the O3-plane interpretation to our orientifold projection.
We also comment on the SUGRA solutions and the duality cascade. 
In section \ref{sec:O3inGenConifold}, we determine the 
orientifold  projection in the generalized conifold. 
Section \ref{sec:Conc} is devoted to conclusion.
 
\section{Preliminary Observation}
\label{sec:Expectation}

\subsection{Expectation from type IIA brane configuration}
\label{sec:IIApicture}
The duality cascade phenomenon \cite{KS} is most easily seen in the type IIA elliptic model
picture.
$N$ D3-branes at the conifold singularity is T-dual to type IIA brane
configurations \cite{DM}: one NS5-brane along the 012345 directions,
the other NS5'-brane along the
012389 directions and $N$ D4-branes along the 01236 directions.
The $x^6$ direction is compactified and four-dimensional 
$\CN=1$ gauge theory is realized on D4-branes along the 0123 directions.

\begin{figure}[htb]
\begin{center}
\leavevmode
\epsfxsize=70mm
\epsfbox{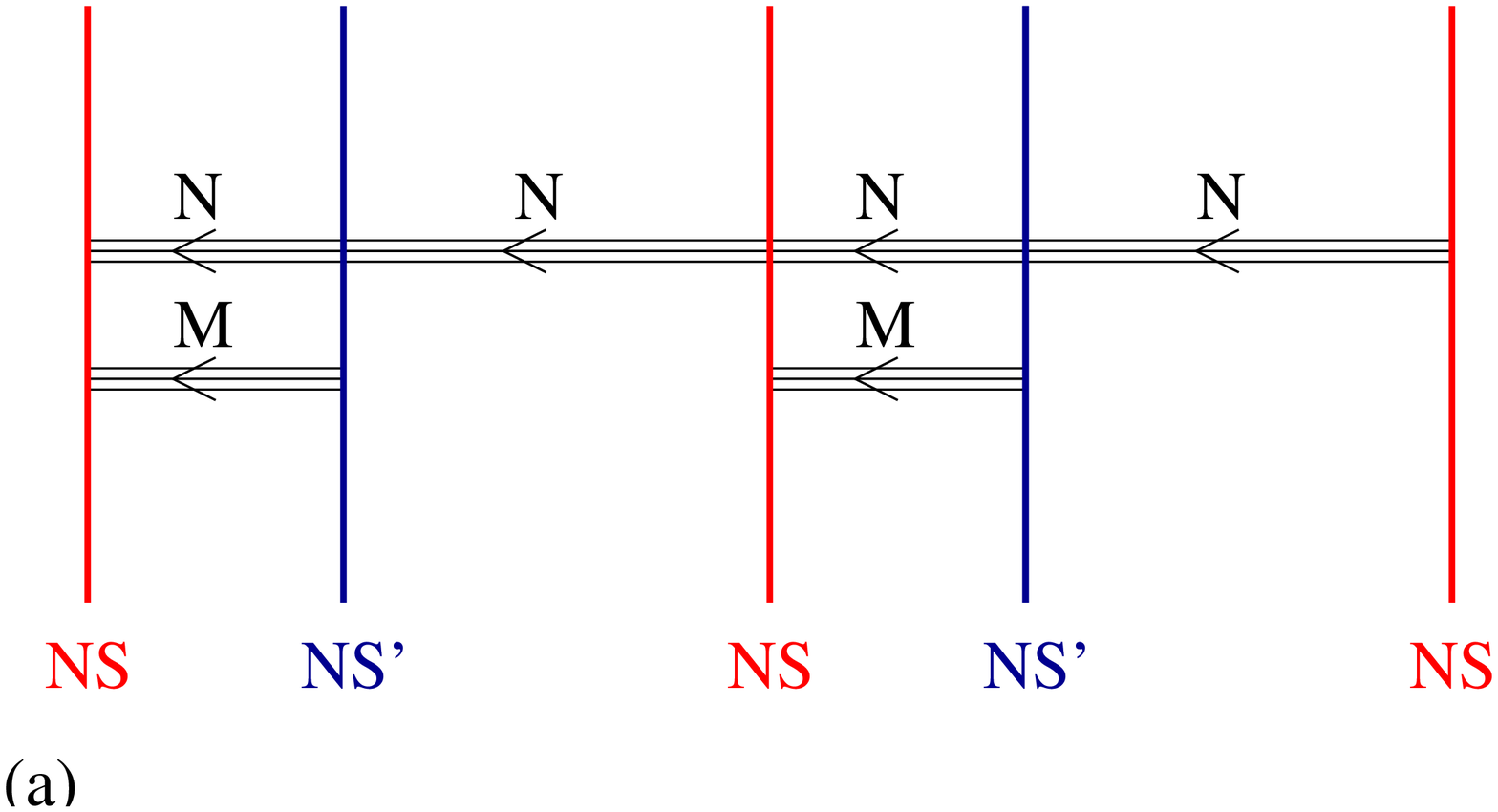}
\hspace{1cm}
\epsfxsize=70mm
\epsfbox{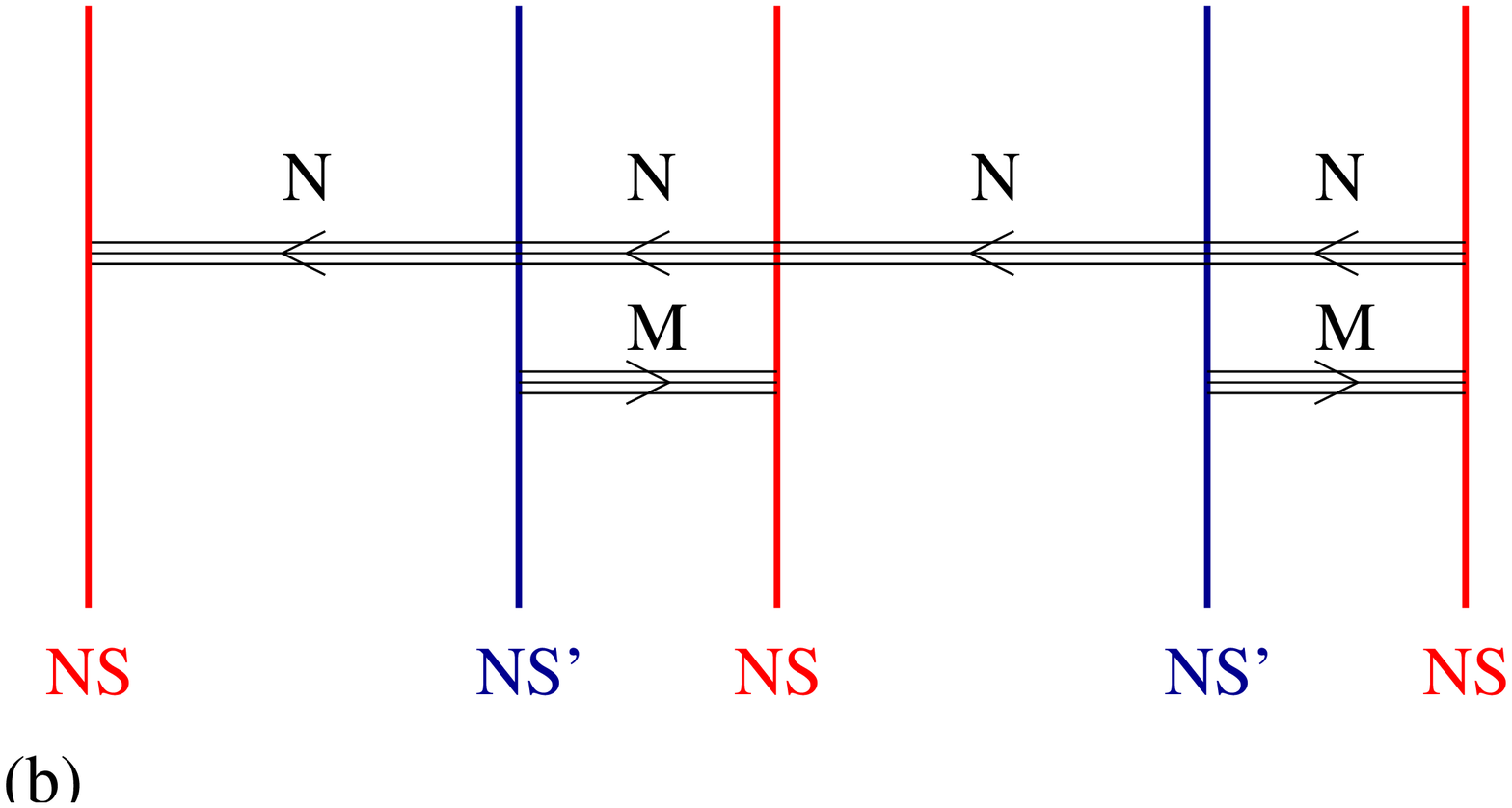}
\caption{
Brane configurations for $SU \times SU$ duality cascade.
The NS5'-brane moves from right to left in this figure.
Fig (a) and (b) are before and after brane crossing.
Two copies of the fundamental region are shown. 
}
\label{fig:SUSU}
\end{center}
\end{figure}

Adding $M$ fractional D3-branes on the type IIB side 
corresponds to adding $M$ D4-branes
stretched between NS5 and NS5' as depicted in  Fig \ref{fig:SUSU} (a).
We call them fractional D4-branes.
The four-dimensional field theory has gauge groups $SU(N+M)\times SU(N)$.
$SU(N+M)$ factor comes from the NS-NS' interval and $SU(N)$ factor comes from the other interval. 
Imbalance of D4-brane tension causes logarithmic bending of NS5-branes
world-volume and positions of
two NS5-branes depend on the energy scale.
This is conveniently described by moving the NS5'-brane.
When the NS5'-brane crosses the NS5-brane, $M$ fractional D4-branes 
in the NS-NS' interval
shrink and re-grow on the other side (Fig \ref{fig:SUSU} (b)). 
This process changes the orientation of fractional D4-branes.
$M$ of $N$ D4-branes are annihilated. 
Then $N-M$ fractional D4-branes in the NS'-NS interval remain.
After all, this brane crossing process changes the gauge groups to
$SU(N)\times SU(N-M)$.

\begin{figure}[htb]
\begin{center}
\leavevmode
\epsfxsize=80mm
\epsfbox{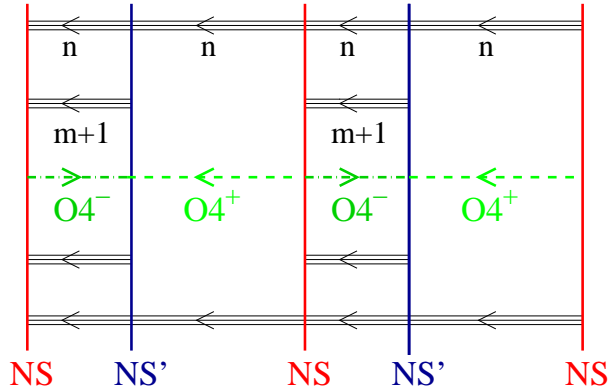}
\caption{
Brane configuration for $SO(N+M+2) \times USp(N)$.
Two copies of fundamental region are shown. 
Here $N=2n,\ M=2m$.
}
\label{fig:SOSP1}
\end{center}
\end{figure}

 $SO\times USp$ gauge theory also exhibits duality cascade phenomenon.
It can be also easily seen in IIA picture.
Let us put the O4-plane on top of D4-branes: the 01236 directions.  
As the O4-plane changes its sign of R-R charge across the NS5-brane \cite{EJS},
 the O4${}^-$-plane in the NS-NS' interval becomes the O4${}^+$-plane
  in the NS'-NS interval. 
When we put $N+M+2$ fractional D4 branes in the NS-NS' interval
and $N$ fractional D4 branes in the NS'-NS interval,
\footnote{
We comment on our convention.
We count the R-R-charges including mirrors. 
For example, O9${}^{-}$ has $-32$ D9-brane charge, and O4${}^{-}$ has
$-1$ D4-charge.
When $M=0$, D4-brane tension between both sides of NS5-branes balances.
}
gauge groups 
 become $SO(N+M+2) \times USp(N)$ (Fig.\ref{fig:SOSP1}).

\begin{figure}[htb]
\begin{center}
\leavevmode
\epsfxsize=80mm
\epsfbox{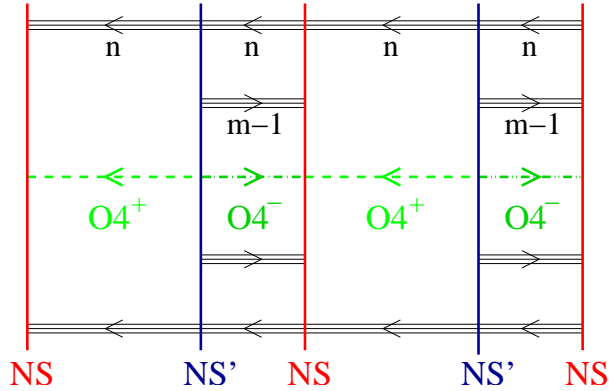}
\caption{
The NS5'-brane has crossed the NS5-brane.
Notice that D4-brane charges flowing away from NS5'-brane is always $M=2m$.
}
\label{fig:SOSP2}
\end{center}
\end{figure}

\begin{figure}[htb]
\begin{center}
\leavevmode
\epsfxsize=80mm
\epsfbox{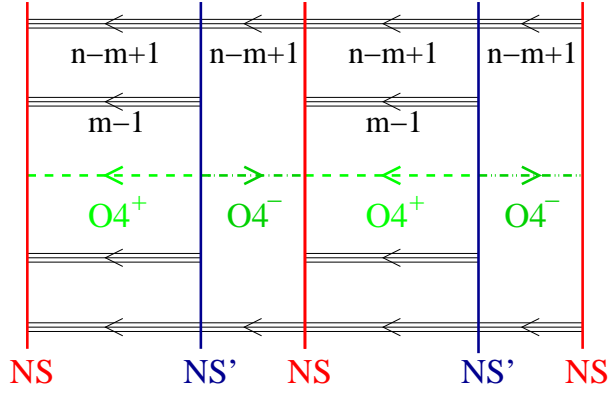}
\caption{
Brane configuration for $USp(N) \times SO(N-M+2)$.
}
\label{fig:SOSP3}
\end{center}
\end{figure}

As opposed to the previous case, when the NS5'-brane crosses the NS5-brane from right to left,
$M+2$ fractional D4-branes shrink and $M-2$ fractional anti-D4-branes emerges
in NS'-NS interval (Fig \ref{fig:SOSP2}). 
The number of D4-branes is determined by
 conservation of D4-brane charge flowing into NS5-branes \cite{HW}.
We must remember that when the O4-plane is crossed by NS5'-brane,
O4${}^+$(O4${}^-$) in the NS'-NS (NS-NS') interval becomes 
O4${}^-$(O4${}^+$) in the NS'-NS (NS-NS') interval respectively.
After pair annihilation process, gauge groups change to 
$USp(N)\times SO(N-M+2)$ (Fig \ref{fig:SOSP3}).

Further brane crossing changes $M-2$ D4-branes in the NS-NS' interval 
into $M+2$ anti-D4-branes in the NS'-NS interval.
After pair annihilation there are $N-M+2$ D4-branes and the O4${}^-$-plane
in the NS-NS' interval, and $N-2M$ D4-branes and the O4${}^+$-plane in
the NS'-NS interval. So the gauge groups become $SO(N-M+2)\times
USp(N-2M)$.

The brane configuration gives us a good understanding for RG cascade,
however, identification of gauge groups and matter contents is rather heuristic.
More detailed discussion is desirable to compare with explicit formulation of the orientifold projection.

\subsection{Expectation from gauge theory}
\label{sec:ExpGaugeTheory}

Before detailed analysis, we comment on the duality cascade 
in terms of field theories.

 Ignoring cumbersome restriction on the rank of gauge groups and number of flavors,
Seiberg dual to $SO(N_c)$ gauge theory with $N_f$ flavors is 
$SO(N_f-N_c+4)$ gauge theory with $N_f$ flavors and singlets \cite{IS}.
And the dual to $USp(N_c)$ gauge theory with  $N_f$ flavors is 
$USp(N_f-N_c-4)$ gauge theory with $N_f$ flavors and singlets \cite{IP}.
\footnote{
For $USp$ gauge theory $N_f$ must be even for the absence of global
anomaly. 
}

 Since $SO(N_1) \times USp(N_2)$ theory is obtained by projection from 
$SU(N_1) \times SU(N_2)$ with $2 (\bN_1,\overline{\bN}_2) \oplus 
2 (\overline{\bN}_1,\bN_2)$, we have $2 (\bN_1,\bN_2)$.
The number of matters are reduced to half compared to $SU \times SU$ theory.
Then duality cascade occurs as following.
\begin{eqnarray}
\label{eq:Cascade}
            &  SO(N+M+2) \times USp(N)& \cr
\Rightarrow &  SO( (N-M)+2) \times USp((N-M)+M)&  \cr
\Rightarrow &  SO( (N-2M) + M + 2) \times USp((N-2M))& \cr
            & \vdots & .
\end{eqnarray}
which is  expected from the type IIA picture.

\section{Determination of Orientifold Projection}
\label{sec:DetOProj}

 In this section, we determine the orientifold projection
in the conifold in terms of  gauge theory on D3-branes.
 From the string theory point of view an orientifold projection is 
product of space-time orbifold projection $R$ and world-sheet
parity $\Omega$ or $\Omega(-)^{F_L}$. 
Because these are symmetries of type IIB string theory,
there exist counterparts in the world-volume gauge theory 
of D3-branes at the conifold singularity.
Luckily, Klebanov and Witten have already identified the space-time symmetry
and $\Omega(-)^{F_L}$ as the global symmetry of the gauge theory \cite{KW}.
We can determine the projection from minor extension of their results.

\subsection{Symmetry of $SU(N_1) \times SU(N_2)$ theory}
\label{sec:SymSUSU}

 The world-volume theory of $N$ D3-branes and $M$ fractional D3-branes on the
conifold singularity is $\CN=1$ supersymmetric $SU(N+M)\times SU(N)$  gauge theory with 
two chiral multiplets $A_1,A_2$ in $(\bf{N+M}, \overline{\bN})$ representation
and two chiral multiplets $B_1,B_2$ in $(\overline{\bf{N+M}}, \bN)$. 
This theory has the superpotential 
\begin{equation}
  W = \lambda \tr (A_i B_j A_k B_l) \epsilon^{ik}\epsilon^{jl}.
\end{equation}
For convenience, we sometimes denote $N_1=N+M$ and $N_2=N$.

 In the following, we briefly review the results on
the dictionary of symmetries of the conifold and gauge theory in the $M=0$ 
case \cite{KW}.
 The moduli space of vacua is the conifold since 
D3-branes can freely move on the conifold.
To see this, suppose that we have diagonal vev of $A_i = a_i := \diag
(a_i^{(1)},...,a_i^{(N)}),B_i = b_i := \diag
(b_i^{(1)},...,b_i^{(N)})$.

Then F-flatness conditions $A_1 B_i A_2=A_2 B_i A_1$ and $B_1 A_i B_2=B_2 A_i B_1$
are trivially satisfied.
Gauge equivalence $a_i \sim e^{i\alpha} a_i, b_i \sim
e^{-i\alpha} b_i$ and D-flatness conditions 
\begin{equation}
  |a^{(r)}_1|^2+|a^{(r)}_2|^2-|b^{(r)}_1|^2-|b^{(r)}_2|^2 =0
\end{equation}
defines the conifold as symplectic quotient.

Another way to see the moduli space as the conifold is
to form gauge invariant quantities $z^{(r)}_{ij} := a^{(r)}_i
b^{(r)}_j $,
 which satisfy the defining equation of the conifold $\det z^{(r)}_{ij} = 0$.
If we denote (we omit the superscript $(r)$ henceforth)
\begin{eqnarray}
\label{eq:z}
z_{ij} = \bpmat z_3 +i z_4 & z_1 -i z_2 \cr z_1 +i z_2 & -z_3 +i z_4
\epmat ,
\end{eqnarray}
the equation is recast into $z_1^2 +z_2^2 +z_3^2 +z_4^2 = 0$.

Symmetries of the $SU(N_1)\times SU(N_2)$ theory is summarized in Table 1,
where $\Lambda_1$ and $\Lambda_2$ are dynamical scale of two gauge groups.
$b$ and $\tilde{b}$ are one-loop beta coefficients and  
$\lambda$ is the coupling constant in the superpotential.

\begin{table}[htbp]
\label{tab:SUSU}
  \begin{center}
    \leavevmode
    \begin{tabular}{|c|cc|ccccc|}
\hline
          &$SU(N_1)$&$SU(N_2)$&$SU(2)$&$SU(2)$& $U(1)_B$  & $U(1)_A$    & $U(1)_R$ \cr
\hline 
     $A_i$ &$\bN_1$&$\overline{\bN}_2$& 2   & 1   &$ 1/2N_1N_2$ & $1/2N_1N_2$ & $1/2$   \cr
     $B_i$ &$\overline{\bN}_1$&$\bN_2$& 1   & 2   &$-1/2N_1N_2$ & $1/2N_1N_2$ & $1/2$    \cr
$\Lambda_1^{b}$&&&  &     &  0        & $2/N_1$     & $2(N_1-N_2)$ \cr
$\Lambda_2^{\tilde{b}}$&&&  &     &  0        & $2/N_2$     &$-2(N_1-N_2)$ \cr
$\lambda$   &&&     &     &  0        & $- 2/N_1N_2$     &  $0$ \cr
$\theta$    &&&     &     &           &           & $-1$      \cr
\hline
    \end{tabular}
  \end{center}
\caption{Quantum numbers of $SU(N_1)\times SU(N_2)$ theory.}
\end{table}

 From the above relation between the conifold and gauge theory, we can
obtain symmetries which are needed for the orientifold projection.
The R-symmetry in gauge theory acts on the fields as 
  \begin{eqnarray}
G_\epsilon\ : \left\{
\bmat    \theta & \mapsto & e^{-i \epsilon} \theta ,\cr
    A_i & \mapsto & e^{\half i \epsilon} A_i , \cr
    B_i & \mapsto & e^{\half i \epsilon} B_i .
\emat \right.
  \end{eqnarray}
We denote the generator of $\BZ_2$ subgroup of this R-symmetry as
 $G(=G_{\pi})$. Althogh $G^2$ changes sign of $A_i,B_i$, this
 is gauge equivalent to $A_i,B_i$. Hence, it is $\BZ_2$ generator.

 From the parameterization $z_{ij} = a_i b_j$,
we can read off transformation rule in SUGRA side 
\begin{eqnarray}
  z_i &\mapsto& e^{i \epsilon} z_i.
\end{eqnarray}
Under this transformation the holomorphic 3-form rotates as
\begin{eqnarray}
 \Omega = \frac{dz_1 \wedge dz_2 \wedge dz_3}{z_4} \mapsto e^{i
 2\epsilon} \Omega .
\end{eqnarray}
Because the holomorphic 3-form can be constructed from covariant constant
spinor as $\Omega_{mnl} = \tp \eta \Gamma_{mnl} \eta $,
it transforms as
\begin{eqnarray}
\eta \mapsto e^{i \epsilon} \eta  .
\end{eqnarray}
Note that covariantly constant spinor and chiral superspace
coordinates rotate oppositely.

 In the same way, the space-time reflection $R_4: z_4 \mapsto -z_4$
 changes the sign of the holomorphic 3-form so covariantly constant spinor
transforms as $\eta \mapsto -i \eta$.
 On the gauge theory side, corresponding $\BZ_2$ transformation becomes
\begin{eqnarray}
\label{eq:R4}
S_1 : \left\{ \bmat
  \CW_1 &\mapsto& \gamma_1 \CW_2 \gamma_1^{-1}  \cr
  \CW_2 &\mapsto& \gamma_2 \CW_1 \gamma_2^{-1}  \cr
  A_i &\mapsto& \epsilon_i{}^j \gamma_1 B_j \gamma_2^{-1}  \cr
  B_i &\mapsto& (\epsilon^{-1})_i{}^j \gamma_2 A_j \gamma_1^{-1} 
\emat \right.
\end{eqnarray}
where $\CW_1$ and $\CW_2$ are field strength multiplets of each gauge group and
$\gamma_1,\gamma_2$ act on Chan-Paton factor to relate D-branes and their mirror images.
$A_i$ and $B_i$ are exchanged because $A_i$ and $B_i$ are spinors of
opposite chirality under $SO(4)\sim SU(2)\times SU(2)/\BZ_2$,
 and reflection $R_4$ acts as gamma matrix $\gamma_4$.
We can replace $\epsilon$ by $g_i{}^j \in SU(2)$ in eq. (\ref{eq:R4}),
and this corresponds to the $SO(3)$ degrees of freedom of $-1$ eigen-vector of reflection.
Note that this is also R-symmetry. Because
superpotential $W$ changes its sign under eq. (\ref{eq:R4}), $\theta$ must rotates 
$\theta \mapsto i\theta$ for superpotential $W$ not to vanish.

Lastly, the world sheet parity $\Omega(-)^{F_L}$ corresponds to 
$\bpmat -1 & \cr & -1 \epmat \in
SL(2,\BZ)$ duality group. 
In particular, it acts 
on unbroken SUSY parameter in the presence of
D3-brane as $\epsilon_L \mapsto \Gamma_{0123} \epsilon_L$.
On the gauge theory side $\Omega(-)^{F_L}$ acts as
\begin{eqnarray}
S_2 : \left\{ \bmat
  \CW_1 &\mapsto& \tp \CW_2 ,\cr
  \CW_2 &\mapsto& \tp \CW_1 ,\cr
  A_i &\mapsto& \tp A_i  ,\cr
  B_i &\mapsto& \tp B_i  .
\emat \right.
\end{eqnarray}
This also changes sign of the superpotential $W$, hence is R-symmetry
: $\theta \mapsto i\theta $.

The dictionary of symmetries on the gauge theory side
and on the type IIB SUGRA side is summarized in Table 2.
\footnote{
In fact, we must pay attention to 
$\pm$ sign in transformation law for spinors .
We use the notation $R_1$ and $\tilde{R}_1$ to distinguish two
  elements of $Spin(6)$ uplifted from $SO(6)$.
For example, $R_{1234}=R_1 R_2 R_3 R_4: \eta \mapsto + \eta$ is different from
  $\tilde{R}_{1234}: \eta \mapsto - \eta$ .
\begin{itemize}
\item The signs for $\theta$ in $S_1,S_2$ may differ relatively.
but it is taken as the same sign in  \cite{KW}.
\item Transformation law for holomorphic 3-form decides that for
  covariant constant spinor $\eta$ only up to sign.
But the sign in $\tilde{R}_{1234}$ is determined by continuity as $U(1)_R$.
\item The sign in $R_4$ is determined by $S_1$ on the gauge theory side.
\end{itemize}
}

\begin{table}
\begin{equation}
\begin{array}{|l|l|}
\hline
{\rm Gauge\ Theory\ side} &{\rm IIB\ SUGRA\ side} \cr
\hline
G \in \BZ_2 \subset  U(1)_R  
&
\tilde{R}_{1234} : {\rm Reflection} \cr
\begin{array}{rl}
\theta \mapsto -\theta  \cr 
A_i \mapsto i A_i \cr
B_i \mapsto i B_i \cr
\end{array}
&
\begin{array}{rcl}
z_\mu & \mapsto &  - z_\mu \cr
\eta & \mapsto & - \eta \cr
\end{array}
\cr
\hline
S_1
& 
R_4 : {\rm Reflection} 
\cr
\begin{array}{rcl}
  \theta &\mapsto& i\theta\cr
  \CW_1 &\mapsto& \gamma_1 \CW_2 \gamma_1^{-1}  \cr
  \CW_2 &\mapsto& \gamma_2 \CW_1 \gamma_2^{-1}  \cr
  A_i &\mapsto& g_i{}^j \gamma_1 B_j \gamma_2^{-1} \cr
  B_i &\mapsto& (g^{-1})_i{}^j \gamma_2 A_j \gamma_1^{-1} \cr
\end{array}
&
\begin{array}{rcl}
(z_1,z_2,z_3,z_4) & \mapsto &  (z_1,z_2,z_3,-z_4) \cr
\eta & \mapsto & -i \eta \cr
\end{array}
\cr
\hline
S_2 &  \Omega (-)^{F_L} : {\rm world \ sheet \ parity} \cr
\begin{array}{rcl}
\theta &\mapsto& i\theta \cr
  \CW_1 &\mapsto& \tp \CW_2 \cr
  \CW_2 &\mapsto& \tp \CW_1 \cr
  A_i &\mapsto& \tp A_i  \cr
  B_i &\mapsto& \tp B_i  \cr
\end{array}
&
\begin{array}{l}
w = \bpmat -1 & \cr & -1 \epmat \in SL(2,\BZ) \cr
\epsilon_L \mapsto \Gamma_{0123} \epsilon_L
\end{array}
\cr
\hline
(SU(2) \times SU(2)) / \BZ_2  & SO(4) \cr
\begin{array}{c}
A_i \mapsto g_i{}^j A_j \cr
B_i \mapsto h_i{}^j B_j \cr
\end{array}
&
z_\mu \mapsto M_\mu{}^\nu z_\nu
\cr
\hline
\end{array}
\nonumber
\end{equation}
\caption{ Correspondence between symmetries of gauge theory side and SUGRA side.}
\end{table}

\subsection{Determination of projection}

Now we can determine the orientifold projection.
Because we expect the resulting theory to posses $\CN=1$ supersymmetry,
the orientifold projection leave the chiral superspace coordinate $\theta$
invariant, otherwise gauge fields and gauginos acquire opposite parity.
Therefore a possible choice for the orientifold projection for this theory 
will be $G S_2 S_1$ .
\begin{eqnarray}
G S_2 S_1 : \left\{ \bmat
  \theta &\mapsto& \theta ,\cr
  \CW_1 & \mapsto & \gamma_1 \tp \CW_1 \gamma_1^{-1} ,\cr
  \CW_2 & \mapsto & \gamma_2 \tp \CW_2 \gamma_2^{-1} ,\cr
  A_i & \mapsto & i g_i{}^j \gamma_1 \tp B_j \gamma_2^{-1} ,\cr
  B_i & \mapsto & i (g^{-1})_i{}^j \gamma_2 \tp A_j \gamma_1^{-1} . 
\emat \right.
\end{eqnarray}
On the IIB SUGRA side, the space-time part of this projection is 
\begin{eqnarray}
R_{123} : (z_1,\ z_2,\ z_3,\ z_4) \mapsto (-z_1,\ -z_2,\ -z_3,\ +z_4).
\end{eqnarray}

In the case of $N_1 \neq N_2$, although we don't know how to separate
world sheet parity $\Omega(-)^{F_L}$ and reflection $R_4$ on the gauge
theory side, it is not necessary for our purpose.
Since $G S_2 S_1$ corresponds to $R_{123} \Omega (-)^{F_L}$ on
the SUGRA side and does not exchange two gauge groups,
we may expect it has the same form as the $N_1 = N_2$ case.

Compatibility of two relations 
$A_i = i g_i{}^j \gamma_1 \tp B_j \gamma_2^{-1}$
and $B_i = i (g^{-1})_i{}^j \gamma_2 \tp A_j \gamma_1^{-1} $ requires
\begin{eqnarray}
  \gamma_1 \tp \gamma_1^{-1} =  - \gamma_2 \tp \gamma_2^{-1} = \pm 1.
\end{eqnarray}
The solution to these conditions is essentially 
\begin{eqnarray}
  \gamma_1 &=& \gamma_{SO} = \bpmat 0 & i {\bf 1} \cr i {\bf 1} & 0 \epmat ,\\
  \gamma_2 &=& \gamma_{Sp} = \bpmat 0 & {\bf 1} \cr - {\bf 1} & 0 \epmat . 
\end{eqnarray}
In particular, combination of $SO$ and $USp$ projection is allowed and agrees 
with the expectation from the type IIA picture.

\subsection{Analysis of the resulting field theory}

 In this subsection, we briefly analyze the field theory after the
orientifold projection in similar manner to the case of $SU\times SU$ gauge theory \cite{KS}.
In the previous section, we have obtained 
\begin{eqnarray}
\label{eq:oproj}
  \CW_1 & \mapsto & \gamma_{SO} \tp \CW_1 \gamma_{SO}^{-1}, \cr
  \CW_2 & \mapsto & \gamma_{Sp} \tp \CW_2 \gamma_{Sp}^{-1}, \cr
  A_i & \mapsto & i g_i{}^j \gamma_{SO} \tp B_j \gamma_{Sp}^{-1}, \cr
  B_i & \mapsto & i (g^{-1})_i{}^j \gamma_{Sp} \tp A_j \gamma_{SO}^{-1}, 
\end{eqnarray}
as the orientifold projection.

This correctly produces $SO(N_1)\times USp(N_2)$ gauge groups,
and matters are reduced to half by the relation $B_i = - i (g^{-1})
\gamma_{Sp} \tp A_i \gamma_{SO}$ 
as expected from the type IIA picture.

The superpotential becomes 
\begin{eqnarray}
\label{eq:SupPot}
  W &=& \lambda \; {\rm tr}(A_i B_j A_k B_l) \epsilon^{ik}
  \epsilon^{jl} \cr
&\sim& \lambda \;  {\rm tr}(A_i \gamma_{Sp} \tp A_m \gamma_{SO}
A_k \gamma_{Sp} \tp A_n \gamma_{SO}) (g^{-1})_j^m (g^{-1})_l^n
\epsilon^{ik}\epsilon^{jl} \cr
&=& \lambda  \; {\rm tr}(A_i \gamma_{Sp} \tp A_j \gamma_{SO} 
A_k \gamma_{Sp} \tp A_l \gamma_{SO}) \epsilon^{ik}\epsilon^{jl} .
\end{eqnarray}

F-flatness conditions are
\begin{equation}
  \tp A_1 \gamma_{SO} A_i \gamma_{Sp} \tp A_2 - 
  \tp A_2 \gamma_{SO} A_i \gamma_{Sp} \tp A_1 =0.
\end{equation}
This can be obtained simply replacing $B_i$ by $-i (g^{-1})_i^j 
\gamma_{Sp} \tp A_i \gamma_{SO}$ in F-flatness conditions of the $SU\times SU$ case.

If we take the vev to be block diagonal form
\begin{eqnarray}
\label{eq:DiagVEV}
 A_i = \bpmat a_i & 0 \cr 0 & \tilde{a}_i \epmat 
&:=& \bpmat
a_i^{(1)} &      &              &   && \cr
          &\ddots&              &   && \cr
          &      & a_i^{(N_2/2)}&   && \cr
{\bf 0} &\cdots&{\bf 0}     &   && \cr
&&&\tilde{a}_i^{(1)} &      &              \cr
&&&          &\ddots&                      \cr
&&&          &      & \tilde{a}_i^{(N_2/2)}\cr
&&&{\bf 0 }&\cdots&{\bf 0}     \cr
\epmat ,\cr  
 B_i = \bpmat b_i & 0 \cr 0 & \tilde{b}_i \epmat 
&:=& \bpmat
b_i^{(1)} &      &              &   && \cr
          &\ddots&              &   && \cr
          &      & b_i^{(N_2/2)}&   && \cr
{\bf 0} &\cdots&{\bf 0}     &   && \cr
&&&\tilde{b}_i^{(1)} &      &              \cr
&&&          &\ddots&                      \cr
&&&          &      & \tilde{b}_i^{(N_2/2)}\cr
&&&{\bf 0 }&\cdots&{\bf 0}     \cr
\epmat ,
\end{eqnarray}
$\tilde{a}_i$ and $\tilde{b}_i$ can be removed due to the projection
(\ref{eq:oproj}). So, we have
\begin{eqnarray}
\label{eq:AfterProj}
  A_i = \bpmat a_i & 0 \cr 0 &  g_i{}^j b_j \epmat ,\ 
  B_i = \bpmat b_i & 0 \cr 0 & - (g^{-1})_i{}^j a_j \epmat. \  
\end{eqnarray}
These automatically satisfy F-flatness conditions. 
In our basis, Cartan subalgebra of both $SO(N_1)$ and $USp(N_2)$ are
$\bpmat X & \cr & -X \epmat$ with $X$ being diagonal.
The D-flatness condition $|a_1|^2+|a_2|^2-|b_1|^2-|b_2|^2=0$
and gauge equivalence $a_i \sim e^{i\epsilon} a_i, b_i \sim
e^{- i\epsilon} b_i$ are reproduced by using only $A_i$.
The moduli space of vacua is still the conifold.

We can form two kinds of meson operators with respect $SO(N_1)$ and
$USp(N_2)$.
\begin{eqnarray}
  M_{ij}^{SO} &:=& \tp A_i \gamma_{SO} (g^{-1})_j{}^l A_l
 = \bpmat & a_i b_j \cr -(g b)_i (g^{-1} a)_j & \epmat ,\cr
  M_{ij}^{Sp} &:=& A_i \gamma_{Sp} (g^{-1})_j{}^l \tp A_l
 = \bpmat & a_i b_j \cr +(g b)_i (g^{-1} a)_j & \epmat .
\end{eqnarray}

Raising the flavor index by $\gamma_{SO}$ or $\gamma_{Sp}$,
we can see these two mesons have the same eigen values,
\begin{equation}
  Z_{ij} \sim M_{ij}^{SO} (-i\gamma_{SO}) \sim M_{ij}^{Sp}
  (\gamma_{Sp}) 
\sim \bpmat a_i b_j & \cr & -(g b)_i (g^{-1} a)_j \epmat .
\end{equation}

Note that the positions of mirror D-branes 
can be read off from the Chan-Paton index structure,
\begin{eqnarray}
\label{eq:MirrorPosition}
  \tilde{a}_i &=& g_i{}^j b_j , \cr
  \tilde{b}_i &=& -(g^{-1})_i{}^j a_j ,\cr
  \tilde{z}_{ij} &=& - g_i{}^k z_{lk} (g^{-1})_j{}^l .
\end{eqnarray}
If we take $g = \epsilon$ in particular,
the effects of the projection to the conifold is
\begin{eqnarray}
R :\  (z_1, z_2 , z_3 , z_4) \mapsto (- z_1, - z_2 , - z_3 , + z_4). 
\end{eqnarray}

\subsubsection{Symmetry}
Chiral operators are also obtained 
from those of the $SU \times SU$ case by replacing $A_i$ with $B_i$,
\begin{equation}
  \CO_n := C_L^{(k_1...k_n)}C_R^{(l_1...l_n)} {\rm Tr}
(A_{k_1} (\gamma_{Sp} \tp A_{l_1} \gamma_{SO})
... A_{k_n} (\gamma_{Sp} \tp A_{l_n} \gamma_{SO})) .
\end{equation}

Global symmetries are reduced to $SU(2),\ U(1)_A$ and $U(1)_R$ which are
summarized in Table 3
\begin{table}[htbp]
  \begin{center}
    \begin{tabular}{|c|cc|ccc|}
\hline
    & $SO(N_1)$ & $USp(N_2)$ &$SU(2)$&$U(1)_A$&$U(1)_R$ \cr
\hline
$A_i$&$\bf{N_1}$&${\bf N_2}$& $2$   &$1/2N_1N_2$ & $1/2$  \cr
$\Lambda_{SO}^{b}$&&         &     & $2/N_1$ & $2 (N_1-N_2-2)$  \cr
$\Lambda_{USp}^{\tilde{b}}$
    &             &         &     & $1/N_2$    &  $-(N_2-N_1+2)$  \cr
$\lambda$
    &             &         &     & $-2/N_1N_2$ & $0$   \cr
$\CO_n$
    &          &         &$(n+1)\otimes(n+1)$&$2n/N_1N_2$  &  $n$   \cr
\hline
    \end{tabular}
    \label{tab:SOSP}
  \end{center}
\caption{Quantum numbers of $SO \times USp$ theory}
\end{table}

The dynamical scale of $USp$ gauge group always appears  through
$\Lambda^{2\tilde{b}}_{Sp}$, hence the anomaly free 
R symmetry is $\BZ_{2M}$ when we take $N_1=N+M+2$ and $N_2=N$.

\subsubsection{RG cascade}

 From the ``Novikov-Shifman-Vainshtein-Zakharov beta function''\cite{NSVZ}, we obtain
\footnote{Here, we have used NSVZ beta function \cite{NSVZ} of the form 
  \begin{eqnarray}
\frac{d}{d \log(\Lambda/\mu)}\frac{8\pi^2}{g^2} = 3T(G) -\sum_i
T(R_i)(1-\gamma_i) \nonumber
  \end{eqnarray}
where $T(R)$ denotes the index of the representation $R$
(it is defined as the normalization of generators $T(R)\delta^{ab} = \tr T^a_R T^b_R $). The summation is taken over
the representation to which the $i$-th matter belongs.

This is not the standard form of the NSVZ beta function.
In duality cascade literature, normalization of the gauge coupling is chosen 
so that denominator of NSVZ beta function is not needed, which is
 commented in \cite{HKO,FKT}.  
The relation with normalization of the gauge coupling and the exact expression of
the beta function was found in \cite{AM}. A simple exposition is given
for example in \cite{A}.  
 }
\begin{eqnarray}
  \frac{d}{d \log(\Lambda/\mu)} \frac{8\pi ^2}{g^2_{SO}}
&=& 3 (N_1 -2) - 2 \times N_2 \times 1 (1- \gamma_A) ,  \\
  \frac{d}{d \log(\Lambda/\mu)} \frac{8\pi ^2}{g^2_{Sp}}
&=& 3 \frac{(N_2 +2)}{2} - 2 \times N_1 \times \half (1- \gamma_A),
\end{eqnarray}
where $\gamma_A$ denotes the anomalous dimension of the matter $A_i$'s.
If we impose conformal invariance, $\gamma_A = -1/2$ and $N_1-N_2 =2$
are required.
These conditions agree with R-R force balance in the type IIA brane configuration picture.
Away from the conformality, $\gamma_A \sim -1/2 +
O(\frac{N_1-N_2}{N_1+N_2})$, we obtain
\begin{eqnarray}
  \frac{d}{d \log(\Lambda/\mu)} \frac{8\pi ^2}{g^2_{SO}}
&\sim& 3 (N_1 - N_2 -2) + O\left(\frac{N_1-N_2}{N_1+N_2}\right),  \\
  \frac{d}{d \log(\Lambda/\mu)} \frac{8\pi ^2}{g^2_{Sp}}
&\sim& 3 \frac{(N_2 - N_1 +2)}{2} + O\left(\frac{N_1-N_2}{N_1+N_2}\right).
\end{eqnarray}
Hence two gauge couplings flow opposite way.

When $N_1 =N+M+2$ and $N_2=N$, the $SO$ gauge group becomes strong
  coupling  and we must perform 
Seiberg duality transformation for reliable description.
We have already verified in sec \ref{sec:ExpGaugeTheory} gauge groups become 
$SO(N-M+2) \times USp(N)$.
Upon this duality transformation, we have
dual quarks $\tilde{A_i}$ and extra singlets 
$M_{ij}^{SO}$ which are mesons of the original theory. 
The superpotential of the dual theory becomes 
\begin{equation}
  W= \lambda  \tr (M_{ij}^{SO} \gamma^{Sp} M_{kl}^{SO} \gamma^{Sp})
\epsilon^{ik}\epsilon^{jl}
+ \frac{1}{\mu} \tr (M_{ij} \gamma^{Sp} 
\tp \tilde{A_k} \gamma^{SO} \tilde{A_l} \gamma_{Sp})
\epsilon^{ik}\epsilon^{jl}.
\end{equation}
Since singlets are massive, we may integrate them out and 
then we have a superpotential of the same form as original theory
(eq. (\ref{eq:SupPot})).

When $N_2 > N_1 - 2$ above analysis applies in the same way.
And we find the $SO \times USp$ duality cascade as in eq. (\ref{eq:Cascade}).

\subsubsection{Deformed conifold as quantum moduli space}

Now we want to show that the quantum moduli space is deformed as in the $SU
\times SU$ case at the bottom of the cascade.
We may suppose $N_1 \gg N_2 \ ( M \gg N ,\ N_1=N+M+2,N_2=N)$ 
or $N_2 \gg N_1 \ (M \gg N , \ N_1=N+2,N_2=N+M)$ as a result
of successive cascade.

Firstly when $N_1 \gg N_2$, $SO(N_1)$ gets strong coupling and
$USp(N_2)$ may be treated as flavor symmetry ( $\Lambda_{Sp}$ can be ignored).
Due to a strong coupling effect Affleck-Dine-Seiberg superpotential 
$W_{ADS} = \left( \frac{\Lambda^b_{SO}}{{\rm Det} M^{SO}_{ij}}
\right)^{1/(N_1-2N_2 -2)}$ is generated, where the determinant is taken 
to $SU(2)$ and $USp(N_2)$ as one flavor index.
Let us take the diagonal form $A_i = \bpmat a_i & \cr & (gb)_i
\epmat$ with first $N_2/2$ nonzero elements $a_i$ taking the same
value and also $b_i$.
Hence the $2N_2$ by $2N_2$ $SO$ meson matrix is brought to $N_2/2$ by $N_2/2$ block diagonal form 
with each diagonal entry as
\begin{eqnarray}
  \bpmat      & z_{11} &      & z_{12} \cr
  (g\tp z \tp g^{-1})_{11} & & (g\tp z \tp g^{-1})_{12} & \cr
        & z_{21} &      & z_{22} \cr
  (g\tp z \tp g^{-1})_{21} & & (g\tp z \tp g^{-1})_{22} & \cr
  \epmat .
\end{eqnarray}
We have $\Det M^{SO}_{ij} = ((\det z_{ij})^2)^{N_2/2} $
, where the small determinant is taken to $SU(2)$ index.
On the other hand $W_{tree} \sim \lambda \tr (M^{SO}_{ij} \gamma_{Sp}
M^{SO}_{kl} \gamma_{Sp}) \epsilon^{ik}\epsilon^{jl} 
\sim \lambda \det z_{ij} $.
The supersymmetric vacuum condition $\del (W_{tree} + W_{ADS}) =0$ is
\begin{eqnarray}
  0 =\left( \lambda - \left[ \frac{\Lambda^{b}_{SO}}{(\det
  z_{ij})^{N_1-N_2-2}} \right]^{1/(N_1-2N_2 -2)} \right) z_{ij}
\end{eqnarray}
Hence, the quantum moduli space becomes the deformed conifold and 
has $M$ branches \ ($N_1=N+M+2,N_2=N$).

Second when $N_2 \gg N_1$, $USp(N_2)$ becomes strong coupling.
Affleck-Dine-Seiberg superpotential is 
$W_{ADS} = \left( \frac{\Lambda^b_{Sp}}{{\rm Pf} M^{Sp}_{ij}}
\right)^{2/(N_2-2N_1 +2)}$
 where Pfaffian is taken to $SU(2)$ and $SO(N_1)$
as one flavor index.
Putting $A_i$ ``diagonal'' as in the $SO$ case,
$USp$ mesons are brought to $N_1/2$ by $N_1/2$ block diagonal form 
with each diagonal entry as
\begin{eqnarray}
  \bpmat      & z_{11} &      & z_{12} \cr
  -(g\tp z \tp g^{-1})_{11} & & -(g\tp z \tp g^{-1})_{12} & \cr
        & z_{21} &      & z_{22} \cr
  -(g\tp z \tp g^{-1})_{21} & & -(g\tp z \tp g^{-1})_{22} & \cr
  \epmat .
\end{eqnarray}
We have $\Pf M^{SO}_{ij} = ((\det z_{ij}))^{N_1/2} $.
The supersymmetric vacuum condition is
\begin{eqnarray}
  0 =\left( \lambda - \left[ \frac{\Lambda^{2\tilde{b}}_{Sp}}{(\det
  z_{ij})^{N_2-N_1+2}} \right]^{1/(N_2-2N_1+2)} \right) z_{ij}
\end{eqnarray}
Again, the quantum moduli space becomes the deformed conifold and 
has $M$ branches \ ($N_1=N+2,N_2=N+M$).

\subsection{Support for our argument}

We give some comments on our orientifold projection.

 Our projection is different from the one proposed by \cite{ANS}. But
 we claim our projection is the correct one for the Klebanov-Strassler 
 model from the following reason. 
The resulting $SO \times USp$ theory must have an $\CN=1$ SUSY as can
 be expected from the IIA picture.
The holomorphic coordinates of the conifold $z_i$ are constructed
 as chiral superfields on the gauge theory side.
Since their projection relates chiral superfield $z_i$ and anti-chiral 
 superfield $\bar{z}_i$, it is not compatible with $\CN=1$
 supersymmetry. 
On the other hand, our projection is determined from $\CN=1$
 supersymmetry as one of the requirement, and 
the resulting field theory exhibits duality cascade. 

As we have noted in the footnote of sec \ref{sec:SymSUSU},
 there are some sign ambiguities in the transformation property of
chiral superspace coordinate $\theta$.
But these ambiguities only affects whether the projection
for $(z_1,z_2,z_3,z_4)$ is $(-,-,-,+)$
or $(+,+,+,-)$.
First projection corresponds to the O3-plane, because the fixed point
 of this projection is located only at the tip of the conifold.
On the other hand, second one has (real) four dimensional fixed point set
in the conifold. Therefore it corresponds to the O7-plane.

 For later convenience, let us relabel coordinates in
eq. (\ref{eq:z}) 
as $x=z_{11},\ y=z_{22},\ z=z_{12},\ w=z_{21}$.
Then our projection acts as $(x,\ y,\ z,\ w) \mapsto (y,\ x,\ ,-z,\
-w)$.
If we take the T-duality, the conifold becomes intersecting NS5-branes
located at $zw=0$. With our choice of coordinates in the IIA picture
(sec \ref{sec:IIApicture}), correspondence of the coordinates 
become $z=x_4+ix_5$ and $w=x_8+i x_9$.
Therefore the projection $(z,w)\mapsto (-z,-w)$ implies that
it gives the O4-plane in the type IIA picture.
At this stage, $(x,y)$ and $(z,w)$ seems to be on equal footing.
So one might suppose that $xy=0$ is also allowed as positions of
intersecting NS5-branes after T-duality.
But as we will see in sec \ref{sec:O3inGenConifold}, 
$(z,w)$ is suitable as the locus of NS5-branes 
when the conifold is viewed as one in the series of the generalized
conifolds.
So our choice of projection is consistently extended to the O4-plane
in the generalized NS5-brane configurations.

 Note that our projection cannot be imposed on the resolved conifold.
The D-flatness condition of the $SU(N) \times SU(N)$ theory is solved
as
\begin{eqnarray}
|A_1|^2+|A_2|^2-|B_1|^2-|B_2|^2   
=
\bpmat
|a_1|^2+|a_2|^2-|b_1|^2-|b_2|^2 & \cr
& |\tilde{a}_1|^2+|\tilde{a}_2|^2-|\tilde{b}_1|^2-|\tilde{b}_2|^2 \cr
\epmat
=\xi {\bf 1}
\end{eqnarray}
where $\xi$ is a constant.
Here we consider the $N_1=N_2$ case because if $N_1 \ne N_2$,
$\xi$ is zero.
In the type IIA picture, $N_1 \ne N_2$ means that $|N_1 -N_2|$
fractional D4-branes are suspended between the NS5-brane and
the NS5'-brane.
Hence it is impossible to pull a single NS5-brane away with keeping
supersymmetry.
Non zero constant $\xi$ means that the conifold singularity is
resolved.
If we introduce the orientifold, $\tilde{a}$ and $\tilde{b}$ relate to 
$b$ and $a$ by eq. (\ref{eq:MirrorPosition}).
Then $\xi$ must vanish.
This is consistent with the type IIA picture
where NS5-branes cannot be pulled away from the O4-plane.

\section{IIB SUGRA solution}
\label{sec:IIBSUGRA}

In this section we investigate the space-time aspects of 
our orientifold projection in more detail and show that KT/KS
solutions \cite{KS,KT} survive with some shifts of boundary
conditions.

\subsection{Fixed points of orientifold projection}
\label{sec:FixedPt}

The singular conifold $z_1^2 + z_2^2 +z_3^2 +z_4^2 =0$ is a cone over 
$T^{1,1}$ space $\sim (SU(2) \times SU(2))/U(1)$. This is easy to 
see in gauge theory or symplectic quotient construction.

Block diagonal elements $a_i,\ b_i$ of chiral superfields $A_i,\ B_i$ is identified as 
$
  a_i \sim e^{i\epsilon} a_i \
$
and
$
   b_i \sim e^{-i\epsilon} b_i.\  
$
We introduce vector and matrix notation, 
${\bf a} := \bpmat a_1 \cr a_2 \epmat,\ 
{\bf b} := \bpmat b_1 \cr b_2 \epmat,\ 
a := \bpmat a_1 & - \overline{a}_2  \cr a_2 & \overline{a}_1 \epmat$
and
$
b := \bpmat b_1 & - \overline{b}_2  \cr b_2 & \overline{b}_1 \epmat
$.
 From D-flatness condition, $|a_1|^2 + |a_2|^2 -|b_1|^2  -|b_2|^2  =
0$,
 the radial coordinate of the cone is defined as $\rho^2 = |{\bf a}|^2 = |{\bf b}|^2$.
When $\rho=1$, $a$ and $b$ belong to  $SU(2)$.

$U(1)$ identification now reads
\begin{eqnarray}
  a \sim  a \bpmat e^{i\epsilon} & \cr & e^{-i\epsilon} \epmat ,\quad
  b \sim  b \bpmat e^{-i\epsilon} & \cr & e^{i\epsilon} \epmat .
\end{eqnarray}

On the other hand, $T^{1,1}$ space is topologically $S^3 \times S^2$.
We can manifestly construct gauge invariant $S^3 \sim SU(2)$ coordinates
\begin{eqnarray}
c = c_\mu \tau_\mu := a \tp b ,
\end{eqnarray}
and $S^2$ coordinate by moment map
\begin{eqnarray}
  n^i := {\bf a}^\dag \tau^i {\bf a}.
\end{eqnarray}

In this notation the orientifold projection is
\begin{equation}
\begin{array}{rcccrcc}
  {\bf a} &\mapsto&    g \, {\bf b}  &, & a & \mapsto & g \ b ,\cr
  {\bf b} &\mapsto&  - g^{-1} {\bf a}&, & b &\mapsto&  - g^{-1} a .
\end{array}
\end{equation}

For $S^3$ coordinates
\begin{eqnarray}
  c=a \tp b &\mapsto& g b \tp (-g^{-1} a) = - g \tp c \tp g^{-1} .
\end{eqnarray}
When $g= \epsilon$, this acts like quaternionic conjugation,
\begin{equation}
\label{eq:Z2onS3}
  (c_1,\ c_2,\ c_3,\ c_4) \mapsto (-c_1,\ -c_2,\ -c_3,\ c_4) .
\end{equation}
For $S^2$ coordinates
\begin{eqnarray}
\label{eq:Z2onS2}
{\bf a}^\dagger \tau^i {\bf a} 
&\mapsto&
 (g \tp c \overline{\bf a})^\dagger \tau^i (g \tp c
  \overline{\bf a}) 
= - {\bf a}^\dagger c \tau^i c^\dagger {\bf a} .
\end{eqnarray}
Here we have used $g=\epsilon$ in the last step.
This is a combined operation of reflection and rotation
that depends on $S^3$ coordinates, $n_i \mapsto -R(c)^i{}_j n^j$. 
 Direct calculation shows that it has eigen value $(-,+,+)$ on the equator
of $S^3\ (c_4=0)$ and $(-,-,-)$ away from the equator.
At the equator $(c_4=0)$, $S^2$ part has fixed point set
$S^1$. 
But the projection acts as anti-podal map on the equator of $S^3$,
so it has no fixed point.
The projection also has no fixed point away from the equator,
 since it act as anti-podal map on both $S^2$ and constant $c_4$
 section of $S^3$.
Therefore, the orientifold projection on the singular conifold
has a fixed point only at the apex, where both $S^2$ and $S^3$
collapse.

\begin{figure}[htb]
\begin{center}
\leavevmode
\epsfxsize=40mm
\epsfbox{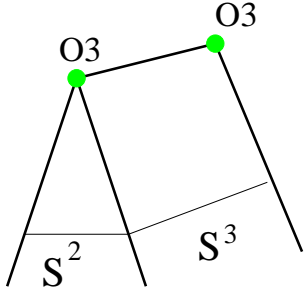}
\caption{
Orientifold 3-plane in deformed conifold.
}
\label{fig:O3onDeformed}
\end{center}
\end{figure}

 Once the action on $S^3 \times S^2$ is known,
we can extend this to the deformed and resolved conifold.
When the conifold is deformed, only $S^2$ shrinks and $S^3$ remains finite
volume at the apex. From eq (\ref{eq:Z2onS3}) it has two fixed points
at the north and south pole of $S^3 (c_4 = \pm 1)$ at the apex of the conifold 
(Fig \ref{fig:O3onDeformed}).
Physically, we have two O3-planes located at the north pole and the south 
pole of the apex. This is in contrast with D3-branes, which are replaced by
fluxes far in the IR of gauge theory \cite{KS,AMV}.
When the conifold is resolved, $S^2$ remains finite volume and $S^3$
shrinks. Since $\BZ_2$ action on $S^2$ depends on $S^3$ coordinates, 
it is not well-defined at the apex. This is also consistent with the
previous analysis.

\subsection{IIB SUGRA solution}

Construction of SUGRA solutions in the presence of orientifold planes
takes two steps.
As we will show, all the fields in the ansatz of KT/KS solutions
have even parity with respect to the orientifold projection $\Omega
(-)^{F_L} R$.
Hence KT/KS solution survives the projection.
All we have to do is simply giving suitable boundary conditions.

In KT/KS solutions, 2-form gauge fields $B_{NS},\ B_{RR}$ and 
their field strengths are
expressed by linear combination of the following 2-forms 
\begin{eqnarray}
\label{eq:2forms}
  &\half (g^1 \wedge g^2 + g^3 \wedge g^4), \\
  & g^1 \wedge g^2 - g^3 \wedge g^4, \\
  & g^1 \wedge g^3 + g^2 \wedge g^4
\end{eqnarray}
and 3-forms
\begin{eqnarray}
\label{eq:3forms}
&\half g^5 \wedge (g^1 \wedge g^2 + g^3 \wedge g^4), \\
 & - g^5 \wedge (g^1 \wedge g^3 + g^2 \wedge g^4), \\
  & g^5 \wedge (g^1 \wedge g^2 - g^3 \wedge g^4). 
\end{eqnarray}
See Appendix \ref{sec:App} for our conventions.
These 2-forms and 3-forms have odd parity under space-time part of the
projection, $R : (z_1, z_2, z_3, z_4) \mapsto (-z_1, -z_2, -z_3, z_4)$.
On the other hand, the metric of the singular/deformed conifold and 
five form field strenth which is proportional to the volume form of $T^{1,1}$ space have even parity under $R$.
As for $\Omega (-)^{F_L}$, 2-forms $B_{NS}$ and $B_{RR}$ have odd parity.
Therefore, all fields in KT/KS solution are even under the whole projection
$\Omega(-)^{F_L} R$.

 Next we consider proper boundary conditions
for the background which corresponds to $SO(N+M+2) \times USp(N)$ gauge theory.
In order to find this, we use the type IIA brane picture. 
In $SU\times SU$ case, whole D4-branes
contribute to D3-charges in type IIB theory 
and fractional D4-branes contribute to D5-charges.
At first sight there might appear $M+2$ fractional D4-branes. 
But two of the fractional D4-branes and the O4${}^-$-plane in the NS-NS' interval 
and the O4${}^+$-plane in the NS'-NS
interval give one unit of D3-charge in type IIB picture.
So D5-charges of this configuration will be $M$!.

In the case of $SO(N+M+2) \times USp(N)$ theory,
we propose the following boundary condition in covering space,
\begin{equation}
\label{eq:BC}
\frac{1}{4\pi^2 \alpha'}  \int_{S^3} F_3 = M, \qquad
\frac{1}{(4\pi^2 \alpha')^2}  \int_{T^{1,1}} F_5 = N+1 .
\end{equation}
The corresponding KT/KS solution is obtained by 
only replacing $N$ with $N+1$.

\subsection{O3${}^+$ or O3${}^-$  ? --- discussion ---}

In the deformed conifold which captures correct IR nature of the gauge 
theory, there are two fixed points at the north and south poles of $S^3$.
Are these fixed points O3${}^-$-plane or O3${}^+$-plane ? 
 From T-dualized type IIA picture,
 we expect that one is O3${}^+$-plane and the other $O3{}^-$-plane.
But boundary conditions in eq. (\ref{eq:BC}) are only aware of overall fluxes and 
seem to smear such microscopic input.

As investigated in \cite{HIS}, the O3${}^+$-plane can be interpreted as 
the O3${}^-$-plane wrapped by an $\BR P^2$-shaped NS5-brane.
So our boundary condition might be understood as O3${}^-$-planes
placed at both north and south poles of $S^3$ and the wrapped NS5-brane is smeared.

\begin{figure}[htb]
\begin{center}
\leavevmode
\epsfxsize=50mm
\epsfbox{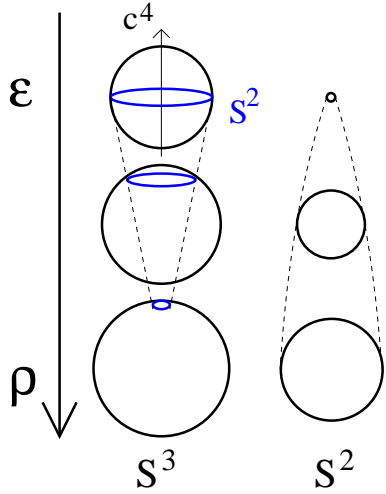}
\caption{
The north pole of $S^3$ at the apex is surrounded by $S^5$ in the conifold.
}
\label{fig:RP5}
\end{center}
\end{figure}

 Aside from the interpretation of our boundary condition,
the orientifolded conifold includes $\BR P^2$ in interesting way.
In the coordinate $c$ and $n^i$, the orbifold part of the orientifold projection
is $(c_1,c_2,c_3,c_4) \mapsto (-c_1,-c_2,-c_3,+c_4)$ and $n^i \mapsto
-R(c)^i{}_j n^j$. Consider $S^2 \times S^2$ obtained by setting $c_4$
to be constant.
Let the $S^2$ in $S^3$ shrink toward the north pole
as we increase radial coordinate from the apex $\rho=\epsilon$ 
(Fig. \ref{fig:RP5}).
In this way, we obtain $S^5$ in the conifold that surround the north
pole ($c_4=1$) as ``join'' of two $S^2$'s.
$\BZ_2$ acts on the $S^5$ as anti-podal map, hence $\BR P^5$ is
obtained.
Direct computation shows that $(n^1,n^2,n^3)=(c_1,c_2,c_3)$ is $-1$ eigen vector
of $-R(c)^i{}_j$. So diagonal $S^2$ becomes $\BR P^2$.
As for the south pole, $\BR P^5$ and $\BR P^2$ are also obtained in the same way.
These two $\BR P^2$ are continuously moved to each other by changing $c_4$.
This implies that configuration of the $O3{}^+$-plane at the north pole and
the $O3{}^-$-plane at the south pole 
is equivalent to the $O3{}^-$-plane at the north pole and the $O3{}^+$-plane at
the south pole.
At the bottom of duality cascade, either of the gauge groups might be
regarded as flavor symmetry, this might be interpreted
 $SO$ and $USp$ gauge theory 
can be continuously interpolated in the IR as found in \cite{AW}.

 The $\BR P^2$ might be essential to explain the duality cascade
phenomenon on the type IIB side. Firstly, let us interpret $N+1$ units of
D3-charge as $N+2-1$. 
$N+2$ of them come from D3-branes and $-1$ comes from two
O3${}^-$-planes.
For convenience sake, let us suppose $N+M+2$ fractional
D3-branes are stuck on the O3${}^-$-plane at the north pole and $N+2$ fractional
D3-branes are stuck on the O3${}^-$-plane at the south pole.
As mentioned above the O3${}^+$-plane can be regarded as the
O3${}^-$-plane wrapped by NS5-brane.  
Two units of fractional D3-charge that convert O3${}^-$-plane to 
O3${}^+$-plane come from Chern-Simons-like coupling on the NS5-brane \cite{HIS}.
Let the NS5-brane enclose the O3${}^-$-plane at the south pole.
Then we may interpret the O3${}^-$-plane and 2 fractional D3-charges 
as the O3${}^+$-plane.
So the gauge group can be identified as $SO(N+M+2)\times USp(N)$.
On each step of duality cascade, D3-charges decrease by $M$ units.
Hence we have $M+(N-M)+2$ fractional D3-charges at the north pole
and $(N-M)+2$ at the south pole. This time we propose that
the NS5-brane encloses the O3${}^-$-plane at the north pole.
So we have $USp(N) \times SO(N-M+2)$ gauge groups
which agree with duality cascade eq. (\ref{eq:Cascade}).
 
 To reproduce the correct cascade, the NS5-brane has to bounce between the
 north and the south pole during the cascade steps.
Above proposal is natural from the view point of the type IIA
 picture, since as mentioned in sec \ref{sec:IIApicture} the
 O4${}^+$-plane in the NS'-NS interval becomes the O4${}^-$-plane
after brane crossing.

\section{Orientifold in Generalized conifolds}
\label{sec:O3inGenConifold}

 In section \ref{sec:DetOProj}, we have succeeded to determine 
the projection on the gauge theory side, which corresponds to 
the O3-plane in the conifold on the type IIB side and the O4-plane in type
IIA elliptic models.
It is tempting to generalize the orientifold projection to the case of
 generalized
conifolds $x y = z^n w^m$.
Again type IIA models are most illustrative pictures where the generalized conifolds are 
realized as transversal $n$ NS5-branes and $m$ NS5'-branes \cite{AKLM}.
In these pictures the O4-plane is still allowed when $n+m$ 
is even \cite{Hori}.

 Above brane configurations are obtained from $N$ parallel
 NS5-branes by rotating some of the NS5-branes.
Since the O4-projection is well-defined through all the 
interpolating theory, it is sufficient to consider the case of
 parallel NS5-branes. This configuration corresponds to $\CN =2,\ \BC^2/\BZ_N \times
 \BC$ quiver gauge theory in the  type IIB picture.

\subsection{Comparison to O6-plane case}

\begin{figure}[htb]
\begin{center}
\leavevmode
\epsfxsize=70mm
\epsfbox{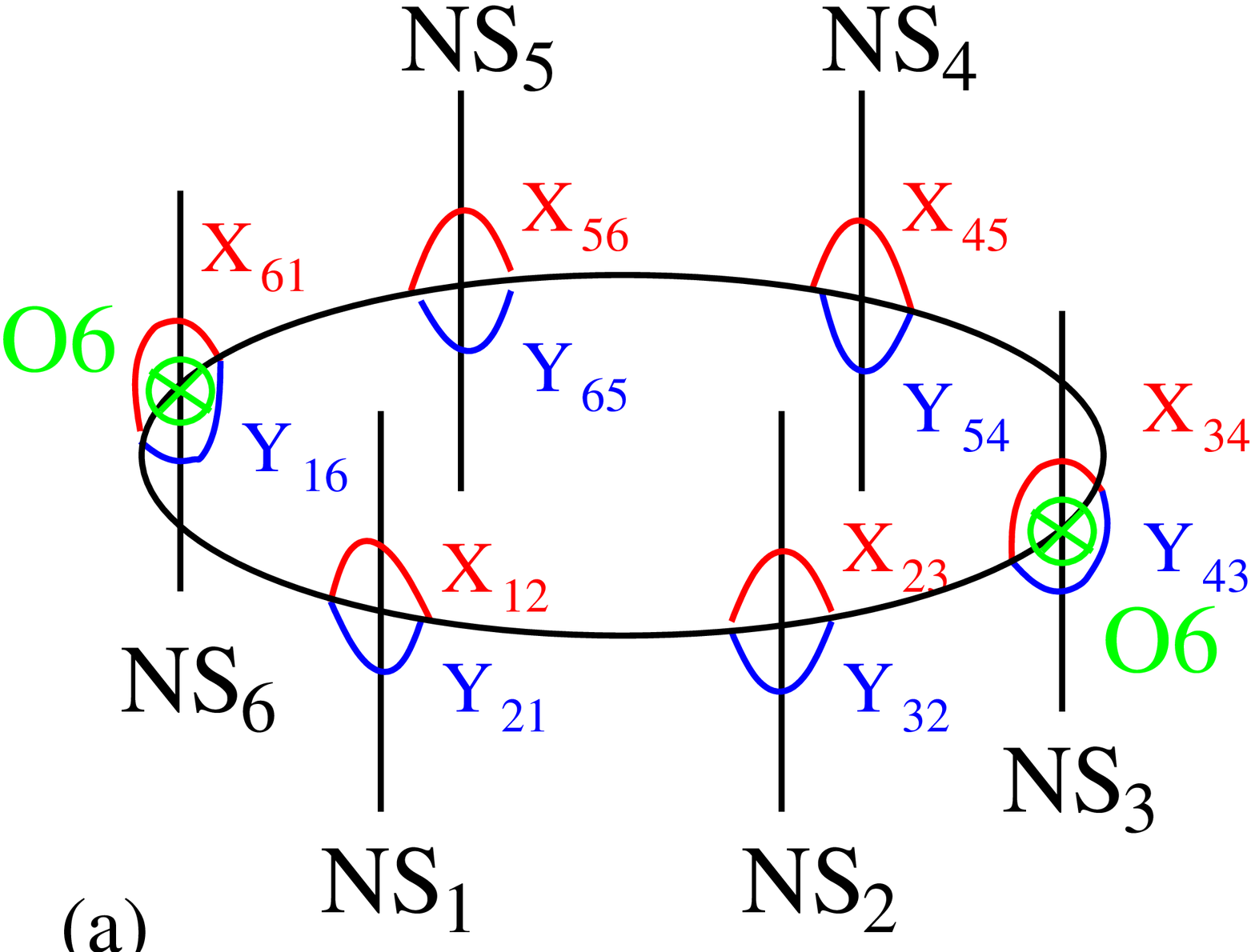}
\hspace{15mm}
\epsfxsize=65mm
\epsfbox{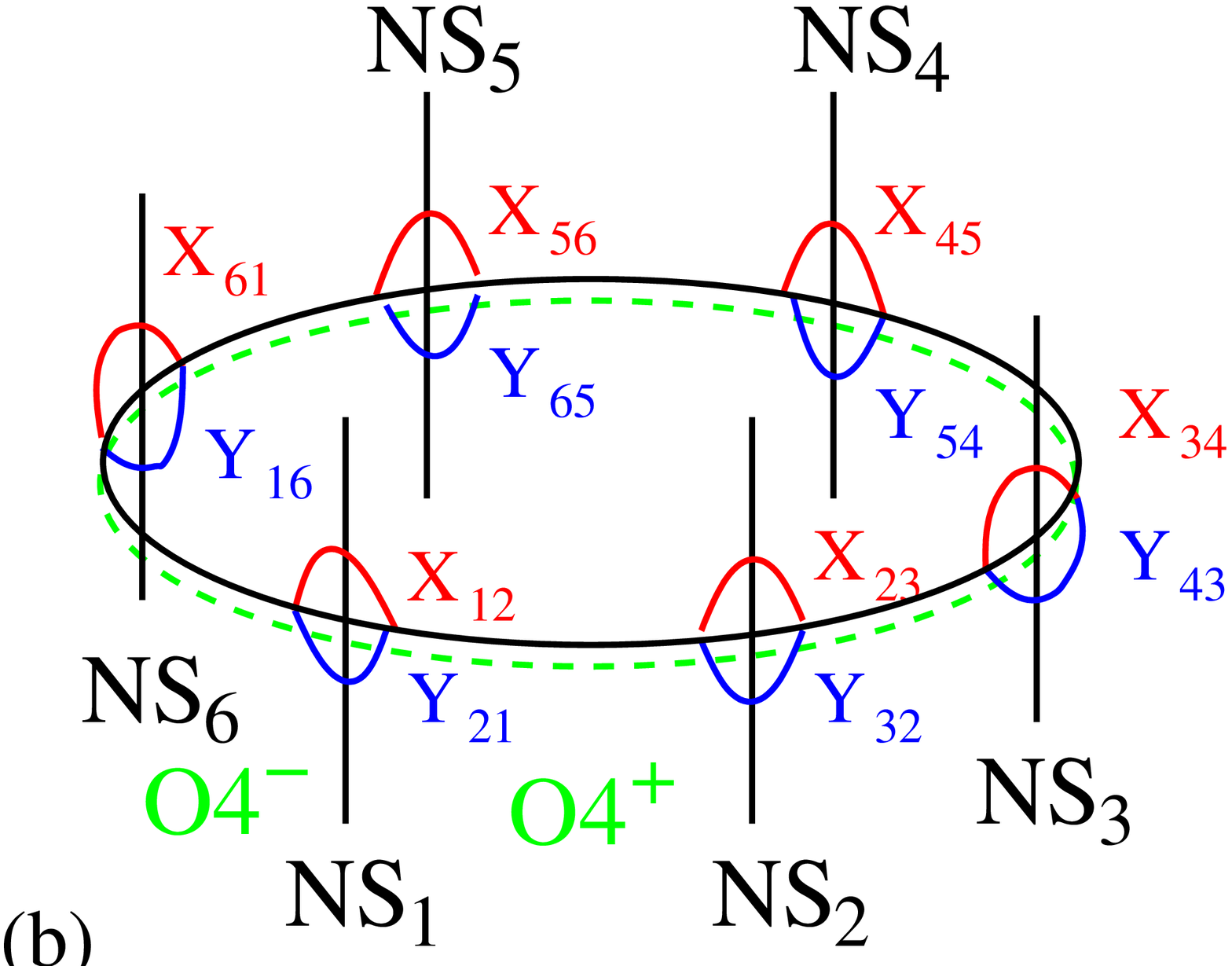}
\caption{
Brane configuration for $SU\times SU \times SU $ model 
and $(SO \times USp)^3$ model.
Open strings connecting $i$-th D4-brane to $(i+1)$-th D4-brane correspond 
to $X_{i,i+1}$ and $Y_{i+1,i}$.
}
\label{fig:O6andO4Model}
\end{center}
\end{figure}

Before analyzing the O4-plane case in the type IIA picture, let us
recall what takes place in the O6-plane case \cite{BI, PU,PRU2}. 
We simply review the $O6^-$ - $O6^-$ case with ${\cal N}=2$ SUSY in four
dimensions \cite{BI, PU}. The brane configurations are $N$ NS5-branes
along $012345$,
D4-branes along $01236$, and 2 $O6^-$-planes along $0123789$.
\footnote{We need D6-branes for the cancellation of tadpoles, however,
they are irrelevant to our discussion. So we omit them.}
We consider the case that two NS5-branes intersect the O6-planes
(Fig. \ref{fig:O6andO4Model} (a)).
Therefore $N$ is even.

Taking the T-duality along the $x^6$ direction, we have D3-branes on the
fixed point of $\BC^2/\BZ_N \times \BC$ singularity with an
O7-plane.
In the first place, let us see the spectrum on the D3-branes
world-volume theory without the O7-plane. 
This theory is obtained from $\CN=4$ $SU$ gauge theory by orbifold 
projection \cite{GP}.
In $\CN=1$ language, this theory has 3 adjoint 
chiral superfields $(X,Y,Z)$ which
correspond to the transverse directions to the D3-branes.
And the superpotential is $W =\tr(Z[X,Y])$.
$\BZ_N$ orbifold projection $\theta:(x,y,z)\mapsto(e^{2\pi i/N}x,e^{-2 
  \pi i/N}y,z)$
 acts on each Chan-Paton sector as
\begin{eqnarray}
  X_{ij} &=& e^{\frac{2\pi i}{N}(i-j+1)} X_{ij} ,\cr
  Y_{ij} &=& e^{\frac{2\pi i}{N}(i-j-1)} Y_{ij} ,\cr
  Z_{ij} &=& e^{\frac{2\pi i}{N}(i-j)} Z_{ij} ,\cr
  \CW_{ij} &=& e^{\frac{2\pi i}{N}(i-j)} \CW_{ij} . 
\end{eqnarray}
Gauge fields surviving the orbifold projection are
\begin{equation}
\CW = 
\bpmat
\CW_{11} &  &      & & \cr
            & \CW_{22} &      & & \cr
            &  &\ddots& & \cr
            &  &      & & \cr
            &  &      &&\CW_{NN}
\epmat
\end{equation}
which give $SU_1 \times SU_2 \times \dots \times SU_N$ gauge groups.
Surviving matter fields are
\begin{equation}
X= 
\bpmat
 & X_{12} &      & & & \cr
 &        &      & & & \cr
 &        &      & \ddots& & \cr
 &        &      & & & \cr
 &        &      & & &X_{N-1,N} \cr
X_{N1}&   &      & & & 
\epmat
,\quad Y =
\bpmat
 & &      & & & Y_{1N}\cr
Y_{21} &        &      & & & \cr
 &        &      &  & & \cr
 &        &      &\ddots & & \cr
 &        &      & & & \cr
 &   &      & &Y_{N,N-1} & 
\epmat
,
\end{equation}
and
\begin{equation}
Z= 
\bpmat
Z_{11} &  &      & & \cr
            & Z_{22} &      & & \cr
            &  &\ddots& & \cr
            &  &      & & \cr
            &  &      &&Z_{NN}
\epmat
.
\end{equation}
$\CW_{i,i}$ and $Z_{i,i}$ are a vectormultiplet of the $i$-th $SU$
gauge group.
$X_{i,i+1}$ and $Y_{i+1,i}$ are $(\bar{\square}_{i},\square_{i+1})$
and $(\square_{i},\bar{\square}_{i+1})$
representation in $SU_{i} \times SU_{i+1}$ groups. Hence they are
combined into  hypermultiplets.

In the type IIA picture, $i$-th gauge group is on $i$-th D4-branes which
suspended between $(i-1)$-th and $i$-th NS5-brane. Therefore
$X_{i,i+1}$ correspond to open strings which run from $i$-th D4-branes
to $(i+1)$-th D4-branes, and 
$Y_{i+1,i}$ are ones from $(i+1)$-th D4-branes to $i$-th D4-branes.

The O7 projections for Chan-Paton matrices \cite{BI,PU} are given by
\begin{equation}
\label{eq:XProj}
X = \gamma_{\Omega'} {}^t X \gamma_{\Omega'}^{-1} 
\end{equation}
where $\Omega'= \Omega (-)^{F_L} R_z$ and
\begin{equation}
\gamma_{\Omega'} =
\bpmat
 & & & & &1\cr
 & & & &\cdots & \cr
 & & &1&      & \cr
 & &-1&       &\cr
 &\cdots& &    & \cr
-1 &     & &   &
\epmat .
\end{equation}
The projections to other fields are similar to $X$.
\footnote{For $\CW$ and $Z$, we need the minus sign in RHS of Eq
(\ref{eq:XProj}).}
These projections give the following relations,
\begin{eqnarray}
\CW_{i,i} &=& - {}^t \CW_{N-i+1, N-i+1}, \\
Z_{i,i} &=& - {}^t Z_{N-i+1, N-i+1}, \\
X_{i,i+1} &=& {}^t X_{N-i, N-i+1} \quad \mbox{for}\quad i\ne
\frac{N}{2}, N,\\
X_{\frac{N}{2},\frac{N}{2}+1} &=&-{}^t
X_{\frac{N}{2},\frac{N}{2}+1},\\
Y_{i+1,i} &=& {}^t Y_{N-i+1, N-i} \quad \mbox{for} \quad i \ne
\frac{N}{2} , N,\\
Y_{\frac{N}{2}+1, \frac{N}{2}} &=& - {}^t Y_{\frac{N}{2}+1,
\frac{N}{2}}.
\end{eqnarray}
Gauge groups $SU_i$ for $N/2<i \le N$ are related to $SU_{N-i+1}$. Hence the
resulting gauge theory is $SU_1 \times SU_2 \times \dots \times
SU_{N/2-1} \times SU_{N/2}$ with matters in 
$\Asym_1 \oplus (\square_1,\square_2)
\oplus (\square_2,\square_3)\oplus \dots\oplus (\square_{N/2-1},\square_{N/2})
\oplus \Asym_{N/2}$ representation.

In the type IIA picture, that orientifold projection nicely matches with
the brane configuration. We take $N/2$-th and $N$-th NS5-branes as
intersecting with O6-planes (Fig. \ref{fig:O6andO4Model} (a)).
 The $i$-th D4-branes are
mirrors to $(N-i+1)$-th D4-branes by the O6-planes. Open strings
corresponding to $X_{i,i+1}$ and $Y_{i+1,i}$ relate to the mirror open
strings $X_{N-i, N-i+1}$ and $Y_{N-i+1, N-i}$ respectively.

The O6-planes relate the D4-branes to ones in the different Chan-Paton 
sector. In the case of the O4-plane, 
the D4-branes are related to ones in the same
Chan-Paton sector (Fig. \ref{fig:O6andO4Model} (b)). So the orientifold projection $\gamma_{\Omega'}$ becomes the
(block) diagonal matrix and relates the open strings to ones in the same
sector as we will see in the next subsection.

\subsection{Determination of the projection in $\BC^2 /\BZ_N \times
 \BC$ case}

 From the type IIA picture, the projection in the type IIB 
$\BC^2 / \BZ_N \times \BC$ 
orbifold relates open strings that
connect adjacent fractional D4 branes with opposit orientation in the same
 Chan-Paton sector.
Therefore, since $X$ and $Y$ are related, 
it is natural to expect the orientifold projection takes following form.
\begin{eqnarray}
  \chi \Omega' : X^I \mapsto 
- \chi^I{}_J  \gamma_{\Omega'} \tp X^J \gamma_{\Omega'}^{-1}.
\end{eqnarray}
where $X^I,\ I=1,2,3$ are $(X,Y,Z)$. 
This projection is combined operation of usual O3 projection
$\Omega'=\Omega(-)^{F_L} R_{XYZ}$ and rotation
$\chi : (X,Y,Z) \mapsto (Y, -X,Z)$. This kind of generalization of the 
O3 projection is allowed in our case
\cite{GP}.

Let us take $\gamma_{\Omega'} = {\rm block.diag}
(\gamma_1,\gamma_2,...,\gamma_N)$.
 From the conditions
\begin{eqnarray}
 X &=& - \gamma_{\Omega'} \tp Y \gamma_{\Omega'}^{-1}\ , \cr
Y &=& \gamma_{\Omega'} \tp X \gamma_{\Omega'}^{-1}\ , \cr
Z &=& - \gamma_{\Omega'} \tp Z \gamma_{\Omega'}^{-1}\ ,
\end{eqnarray}
we obtain the following relations,
\begin{eqnarray}
\label{eq:XY}
\left\{ 
\bmat
  X_{12} &=& - \gamma_{1} \tp Y_{21} \gamma_2^{-1} \cr
&&\vdots \cr
 X_{i,i+1} &=& - \gamma_{i} \tp Y_{i+1,i} \gamma_{i+1}^{-1} \cr
&&\vdots \cr
 X_{N1} &=& - \gamma_{N} \tp Y_{1N} \gamma_1^{-1} \ ,
\emat
\right. 
\end{eqnarray}
\begin{eqnarray}
\label{eq:YX}
\left\{ 
\bmat
  Y_{21} &=& \gamma_{2} \tp X_{12} \gamma_1^{-1} \cr
&& \vdots\cr
 Y_{i+1,i} &=& \gamma_{i+1} \tp X_{i,i+1} \gamma_i^{-1} \cr
&& \vdots\cr
 Y_{1N} &=& \gamma_{1} \tp X_{N1} \gamma_N^{-1} \ ,
\emat
\right. 
\end{eqnarray}
\begin{eqnarray}
\label{eq:ZZ}
\left\{ 
\bmat
  Z_{11} &=& -\gamma_{1} \tp Z_{11} \gamma_1^{-1} \cr
&& \vdots \cr
 Z_{i,i} &=& -\gamma_{i} \tp Z_{i,i} \gamma_i^{-1} \cr
&& \vdots \cr
 Z_{NN} &=& -\gamma_{N} \tp Z_{NN} \gamma_N^{-1} \ .
\emat
\right. 
\end{eqnarray}
First relations eq. (\ref{eq:XY}) are rewritten as 
$Y_{i+1,i} =  - \tp \gamma_{i+1} \tp X_{i,i+1} \tp \gamma_i^{-1}$.
Therefore compatibility condition with second relations eq. (\ref{eq:YX}) requires
\begin{eqnarray}
\label{eq:Compatibility}
\gamma_{i} \tp \gamma_{i}^{-1} = - \gamma_{i+1} \tp \gamma_{i+1}^{-1} 
= \pm {\bf 1} .
\end{eqnarray}
Taking $\gamma = {\rm block.diag.}
(\gamma_{SO},\gamma_{Sp},...\gamma_{SO},\gamma_{Sp})$, the relations in eqs. (\ref{eq:ZZ}) become
\begin{eqnarray}
  Z_{ii} =
\left\{
\bmat
 - \tp \gamma_{SO} \tp Z_{ii} \gamma_{SO}^{-1} \qquad i:{\rm odd} \cr
 + \tp \gamma_{Sp} \tp Z_{ii} \gamma_{Sp}^{-1} \qquad i:{\rm even}
\emat
\right.
\end{eqnarray}
which give adjoint matters of $\CN=2$ vectormultiplets.
The relation eq. (\ref{eq:Compatibility}) restricts $N$ to be even. Moreover gauge groups have alternating
structure
$SO \times USp \times SO \times USp \times ... \times SO \times USp$.
These are specific to the O4-plane in the type IIA picture \cite{Lopez,Hori}!
So we have the O3-plane in the orbifold theory. 
Our projection is consistent with the results found by \cite{Uranga2}
in the context of O5-D5 systems. 
We take N even below.

To sum up the field theory becomes $\CN=2 \ SO_1\times USp_2 \times
...\times SO_{N-1} \times USp_N $ gauge theory with matters in $
\oplus_i^N ( \square_i ,\square_{i+1})$ representation.
 
 We give some remarks.
Combined with $\BZ_N$ orbifold group,
the orientifold projection group has the structure
$\BZ_N \oplus \BZ_N \chi \Omega'$.
$\chi \Omega' $ is required to be order 4. 
We can verify it explicitly
\begin{eqnarray}
  (\chi \Omega')^4 : X^I :(-)^4 (\chi^4)^I{}_J (\gamma \tp
  \gamma^{-1})^2 X^J (\gamma \tp \gamma^{-1})^{-2}.
\end{eqnarray}
Since our solution satisfies $(\gamma \tp \gamma^{-1}=
\bdiag({\bf 1,-1,1,-1,...,1,-1}))$, it has exactly order 4.

\subsection{Generalized conifold case}

As remarked before, once the orientifold projection is obtained in the 
orbifold $\BC^2/\BZ_N \times \BC$, we can extend this result to the 
generalized conifold case.

The orbifold $xy= w^N$ is deformed to 
the generalized conifold $xy =z^n w^m$ $(n+m=N)$ by complex structure deformation
with the interpolating equation $xy = \prod_{i=1}^N (w - \alpha_i z)$,
where $\alpha_i$'s are deformation parameters.
In terms of type IIA theory, this corresponds to arbitrary rotated
NS5-branes.
In the gauge theory, this corresponds to mass deformation
$W_m = \tr ( M Z^2)$ where $M$ is a certain mass matrix.

To investigate the effect of the orientifold projection on the
space-time,
we see the relation between the coordinates $(x,y,z,w)$ and
matter fields $(X,Y,Z)$.
The moduli space of vacua of the $\CN=2$ gauge theory can be identified 
as $xy=w^N$ as follows.
For this purpose, it is sufficient to assume that each field 
has diagonal expectation values.

The F-flatness condition $[X,Z]=0$ requires $Z_{i,i} X_{i,i+1} = X_{i,i+1} 
Z_{i+1,i+1}$, hence the solution is $Z_{11}=Z_{22}=...=Z_{NN}$.
$[X,Y]=0$ requires $X_{i,i+1}Y_{i+1,i}=Y_{i,i-1}X_{i-1,i}$, hence
$X_{12}Y_{21}=X_{23}Y_{32}=...=X_{N1}Y_{1N}$.
$[Y,Z]=0$ requires no further constraint.
Then gauge invariant operators modulo F-flatness conditions are
\begin{eqnarray}
  x &=& X_{12} X_{23} \cdots X_{N1} ,\cr
  y &=& Y_{21} Y_{32} \cdots Y_{1N} ,\cr
  w &=& X_{12}Y_{21} = X_{23}Y_{32} = \dots = X_{N1}Y_{1N} ,\cr
  z &=& Z_{11}= Z_{22} = \dots = Z_{NN}.
\end{eqnarray}
These operators obey one constraint $xy = w^N $ which is the 
defining equation of
$\BC^2/\BZ_N$ as hypersurface and $z$ parameterizes a complex plane $\BC$.

Our projection eqs (\ref{eq:XY}), (\ref{eq:YX}) and (\ref{eq:ZZ}) acts
on $x,y,z,w$ as
\begin{eqnarray}
(x,y,z,w) & \mapsto & (y,x,-z,-w).
\end{eqnarray}
Here we used the fact that $N$ is even.
In our projection $z$ and $w$ have the same parity.
Therefore we can deform the orbifold $xy=w^N$ to the generalized
conifolds $xy=z^m w^n$ through $xy = \prod_{i=1}^N (w-\alpha_i z)$
under our projection.
This is consistent with the type IIA picture in which NS5-branes can be
freely rotated.

This is in contrast to the O6-projections in \cite{PRU2}, 
in which $z$ and $w$ have opposite parity.
So the defining equation is deformed only pairwise 
$xy =\prod_{i=1}^{N/2}(z-\alpha_i w)(z+\alpha_i w)$.

\section{Conclusion}
\label{sec:Conc}
We have determined the orientifold projection in the conifold in type IIB theory. 
This has been obtained by analysis of the correspondence between symmetries of 
the field theory realized on the world-volume of D3-branes and type IIB SUGRA following 
Klebanov and Witten \cite{KW}.

In the type IIB SUGRA picture, the spacetime reflection of the orientifold projection
maps the coordinates of the conifold $(z_1,z_2,z_3,z_4) $ to $(-z_1,-z_2,-z_3,z_4)$.
This orientifold projection has been identified as the O3-plane. In the T-dualized type IIA theory, 
this becomes the O4-plane. Our orientifold projection freezes the parameter of blowing up the 
singularity of the conifold since FI-parameter must vanish under the projection.
This is consistent with the type IIA picture where we can not pull 
a single NS5-brane away
from the O4-plane.

In terms of the field theory, the projection reduces the $SU \times SU$ gauge theory
to the $SO \times USp$ gauge theory.
 From the field theory analysis, we have found that duality cascade phenomenon occurs in 
RG flow like $SU \times SU$ theory \cite{KS}.
This has been also expected from the type IIA brane configuration.
At the bottom of the cascade the singularity of the conifold is
deformed by Affleck-Dine-Seiberg superpotential as in the $SU \times SU$ case again.

We have also found that the corresponding SUGRA solutions to the $SO
\times USp$ gauge theory can be obtained by only modifying the boundary
conditions for the R-R-charges in KS and KT solutions \cite{KS,KT}.
This is better understood in type IIA picture,
since if we focus on the R-R-charges the combination of O4${}^+$, O4${}^-$-planes 
and two fractional D4-branes gives one whole D4-brane charge. The boundary 
condition has matched with the duality cascade phenomenon.

We have extended the orientifold projection to the case of 
generalized conifolds. 
The projection 
requires the total number of NS and NS'-branes to be even. Moreover the 
gauge groups become $SO \times USp \times \cdots \times SO \times USp$.
These properties are consistent  with the features of the O4-plane \cite{Lopez, Hori}.
The projection agrees with one which we have obtained by analysis to the conifold.
 
\section*{Acknowledgments}
We would like to thank Kazutoshi~Ohta for the collaboration in the early
stage of this work and instructive advice.
S.~I. thanks Y.~Hyakutake for helpful comments on interpretation
of the O3-plane in our projection. 
We also thank I.~Kishimoto for helpful discussion and the referee of
Physical Review D for careful reading of our manuscript and comments.

The work of T.\,Y.\ is supported in part by the JSPS Research Fellowships

\appendix
\vspace{1.5cm}
\centerline{\Large\bf Appendix}
\appendix

\section{Orientifold projection on popular parameterization}
\label{sec:App}

The conifold and deformed conifold metrics are given in \cite{KS,CO}.
In the literature $Z=z_\mu \tau_\mu$ is often
parameterized by two $SU(2)$ matrices 
\begin{eqnarray}
L_i = \bpmat 
\cos\frac{\theta_i}{2}e^{i(\psi_i+\phi_i)/2} &
-\sin\frac{\theta_i}{2}e^{-i(\psi_i-\phi_i)/2} \cr
\sin\frac{\theta_i}{2}e^{i(\psi_i-\phi_i)/2} &
\cos\frac{\theta_i}{2}e^{-i(\psi_i+\phi_i)/2} \epmat ,
\end{eqnarray}
as
\begin{eqnarray}
  Z = L_1 Z_0 L_2^\dagger,
\end{eqnarray}
where
\begin{eqnarray}
  Z_0 = \left\{ 
\bmat
\bpmat &1\cr & \epmat & {\rm for \ the \ singular\ conifold} ,\cr
\bpmat & \epsilon e^{\tau/2}\cr \epsilon e^{-\tau/2} & \epmat & {\rm
for \ the \ deformed\ conifold} .
\emat
\right. .
\end{eqnarray}

Basis of 1-forms on $T^{1,1}$ are
\begin{eqnarray}
&&  g^1 =\frac{1}{\sqrt{2}}(e^1-e^3),\   g^2 =\frac{1}{\sqrt{2}}(e^2-e^4), \cr
&&  g^3 =\frac{1}{\sqrt{2}}(e^1+e^3),\   g^4 =\frac{1}{\sqrt{2}}(e^2+e^4), \cr
&& g^5 =e^5 
\end{eqnarray}
where
\begin{eqnarray}
&& e^1    := -\sin\theta_1 d \phi_1,\ \  e^2 := d\theta_1, \cr
&& e^3    := \cos\psi \sin\theta_2 d\phi_2 -\sin\psi d\theta_2, \cr
&& e^4    := \sin\psi \sin\theta_2 d\phi_2 +\cos\psi d\theta_2, \cr
&& e^5    := d\psi + \cos \theta_1 d\phi_1 +\cos \theta_2 d\phi_2.
\end{eqnarray}
Here $\psi_1$ and $\psi_2 $ appear only through the
 combination $\psi=\psi_1+\psi_2$
which has period $4\pi$.

Once the orientifold projection written in $z_{ij}$, this can be 
also used for the projection on the deformed conifold.
In fact, there are two-ways to write space-time $\BZ_2$ in terms 
of $L_i$'s. But this ambiguity is artifact of $U(1)$ redundancy.
A convenient choice will be
$L_1 \mapsto g \bar{L}_2 i \tau_1,\ L_2 \mapsto - \tp g \bar{L}_1 i
\tau_1 $.
When $g=\epsilon$, it is written in angular coordinate as
\begin{eqnarray}
  \theta_1 \leftrightarrow \theta_2,\   \phi_1 \leftrightarrow
  \phi_2,\
\psi \mapsto \psi + 2\pi
\end{eqnarray}
One-forms transform
\begin{eqnarray}
  \bpmat e_1 \cr e_2 \epmat \mapsto 
\bpmat -\cos\psi & -\sin\psi \cr
- \sin\psi & \cos\psi \epmat \bpmat e_3 \cr e_4 \epmat,\ \
  \bpmat e_3 \cr e_4 \epmat \mapsto 
\bpmat -\cos\psi & -\sin\psi \cr
- \sin\psi & \cos\psi \epmat \bpmat e_1 \cr e_2 \epmat .\ 
\end{eqnarray}
It is easy to verify various 2-forms and 3-forms in eqs. (\ref{eq:2forms}),(\ref{eq:3forms})
is odd under the space-time part of the projection.
It can be more easily verified $SO(4)$ invariant expression found in
\cite{HKO}. 

 Note that $SU(2)\times SU(2)$ to which $(L_1,L_2)$ belong 
is different from $SU(2) \times SU(2)$ to which $(a,b)$ belong in
section \ref{sec:FixedPt}.

\end{document}